\documentclass[a4paper]{jpconf}
\usepackage{graphicx}
\usepackage{amsmath,amsfonts,amsthm}

\begin{document}
\title{How vacuum fluctuations determine the properties of the vacuum}

\author{G. B. Mainland$^1$ and Bernard Mulligan$^2$}

\address{$^1$ Department of Physics, The Ohio State University at Newark, 1179 University Dr., Newark, OH 43055, USA, $^2$ Department of Physics, The Ohio State University, Columbus, OH 43210, USA}

\ead{mainland.1@osu.edu, mulligan.3@osu.edu}

\begin{abstract}
Particle-antiparticle pairs are predicted by quantum field theory to appear as vacuum fluctuations. The model of the vacuum used here is postulated to have the following properties:  To minimize the violation of conservation energy allowed by the Heisenberg uncertainty principle and to avoid violating conservation of angular momentum,  vacuum fluctuations of charged particle-antiparticle pairs appear as bound states in the lowest energy level that has zero angular momentum. These transient atoms  are polarized by electric fields somewhat similarly to the way that ordinary matter is polarized.  As a consequence, the permittivity $\epsilon_0$ of the vacuum can  be calculated.  Once the permittivity of the vacuum has been calculated, formulas for the speed of light $c$  in the vacuum and the fine-structure constant $\alpha$  immediately follow. The values  for $\epsilon_0$, $c$, and $\alpha$ calculated here agree with the accepted values to within a few percent. Only the leading terms in the formulas have been retained in the calculations. The absence of dispersion in the vacuum is discussed and explained.
\end{abstract}

\section{Introduction: vacuum fluctuations \textendash \,  the vacuum as a dielectric}
\label{sec:1}

Values of the permittivity $\epsilon_0$ of the vacuum and the speed $c$ of light in the vacuum are properties of the vacuum. Initially it is not obvious that the value of the fine-structure constant $\alpha$ also is a property of the vacuum. However, as will be shown later in the Introduction, if one of the three quantities $\epsilon_0, \,c,$ or $\alpha$ has been calculated, values for the other two immediately follow. Consequently, the value of $\alpha$ must also be a property of the vacuum. 

Among the three quantities $\epsilon_0, \,c,$ and $\alpha$, physicists  have had little idea how to calculate either $c$ or $\alpha$.  But for almost a century they have known how to calculate the permittivity of a dielectric\cite{Kramers:24}.  As a consequence, most of this article is devoted to calculating $\epsilon_0$, guided along the way by the well-established procedures for calculating the permittivity of dielectrics.  Once $\epsilon_0$ has been calculated, values for $c$ and $\alpha$ immediately follow.

Vacuum fluctuations (VFs) are  particle-antiparticle pairs that appear spontaneously in the vacuum as predicted by relativistic quantum field theory\cite{Thirring:58,Bjorken:65,Roman:69}, violating conservation of energy to the extent allowed by the Heisenberg uncertainty principle.  VFs are created on-mass-shell and are not associated with a Green's function or perturbation theory.

At this point a word of caution is required:   Vacuum bubbles\cite{Bjorken:65,Peskin:95} are a class of perturbation effects in quantum field theories that are sometimes confused with VFs because their Feynman diagrams look like bubbles that originate from and terminate in the vacuum. Vacuum bubbles are not VFs and, in fact, do not make a contribution to physical processes\cite{Bjorken:65,Peskin:95}.

The idea that  VFs  play a role in determining the permittivity of the vacuum is very old:  in 1934 Furry and Oppenheimer\cite{Furry:34} wrote that VFs of charged particle-antiparticle pairs would affect the value of the dielectric constant of the vacuum; ``Because of the polarizability of the nascent pairs, the dielectric constant of space into which no matter has been introduced differs from that of truly empty space.''  In 1936 the idea of treating  the vacuum as a medium with electric and magnetic polarizability was discussed by  Weisskopf and Pauli\cite{Weisskopf:94,Pauli:94}.  

For VFs to be responsible for the permittivity of the vacuum, it is crucial that they possess elasticity (an effective spring constant). The possibility that a charged lepton-antilepton pair can form an atomic bound state, which possesses precisely this property, was discussed by Ruark\cite{Ruark:45}.  At almost the same time, the formation of a charged lepton-antilepton bound state was elaborated on by Wheeler\cite{Wheeler:46}  when he calculated the decay rate of parapositronium, the spin-0, ground state of a bound electron and positron, into two photons. The experimental proof that such atoms exist was provided by Deutsch and Brown\cite{Deutsch:52}. 

In 1957 Dicke\cite{Dicke:57} wrote about the possibility that the vacuum could be considered as  a dielectric medium.  More recently the possibility that the properties of the quantum vacuum determine, in the vacuum, the speed of light and the permittivity has been explored by a number of authors\cite{Leuchs:10,Leuchs:13,Urban:13}. 
 
As will be shown, the VFs that primarily contribute to the permittivity of the vacuum are charged lepton-antilepton pairs.  To conserve angular momentum and minimize the violation of conservation of energy,  charged lepton-antilepton VFs must appear in the vacuum as atoms.  The term ``dielectric'' can then be used in the usual sense:  a photon passing through  the vacuum is slowed by its interactions with transient atoms, a concept familiar from discussions of a physical dielectric\cite{Kramers:24,Rossi:59,Feynman:63}. 

In the discussion that follows, a quantum calculation of the the permittivity of the vacuum is presented that is similar both to a calculation of the permittivity of a physical dielectric\cite{ Sokolov:84} and to a previous quantum calculation of the the permittivity\cite{Mainland:18a} of the vacuum by the authors. The result of the calculation is \footnote{SI units are used throughout.}
\begin{equation}\label{eqn:1}
\epsilon_0 \cong   \frac{ 6\mu_0}{\pi}\left(\frac{8e^2}{\hbar}\right)^2= 9.10\times 10^{-12}\rm \frac{C}{Vm}\, .
\end{equation}
In the above  equation  $e$ is the magnitude of the charge on an electron,  $\hbar$ is Planck's constant divided by $2\pi$, and $\mu_0$ is the permeability of the vacuum with an assigned value $4\pi \times 10^{-7}$H/m.  Thus in  Equation  \eqref{eqn:1}, $\epsilon_0$ is expressed in terms of two experimentally-determined quantities, $e$ and  $\hbar$. The experimental value for $\epsilon_0$ is  2.8 \% less than the calculated value. The calculation of $\epsilon_0$  is  simplified, and the numerical accuracy is reduced, by including  (a) only contributions to lowest order in what turns out to be an expansion in powers of the fine-structure constant $\alpha$ and (b) only the interactions of photons with bound states  of charged lepton-antilepton VFs. 

Once $\epsilon_0$ has been calculated, formulas for $c$ and  $\alpha$ immediately follow. Using $c=1/\sqrt{\mu_0\epsilon_0}$ and the formula for $\epsilon_0$ in    Equation \eqref{eqn:1},
\begin{equation}\label{eqn:2}
c \cong   \sqrt{\frac{\pi}{6}}\frac{\hbar}{8e^2\mu_0}= 2.96\times 10^8\rm{m/s}\,.
\end{equation}
The accepted value is $c=3.00 \times 10^8$m/s, which is 1.3\% more  than the calculated value. The progress of a photon traveling through the vacuum is slowed when it interacts with and has a polarizing effect on a VF consisting of a charged lepton and antilepton bound into an atom, implying that the interaction of photons with vacuum fluctuations determines the speed of light in the vacuum. 

Any observer in an inertial frame of reference cannot detect relative motion of the vacuum and the inertial frame. Thus the observer would conclude that the vacuum is at rest with respect to the inertial frame. Leonhardt et al.\cite{Leonhardt:18}  state, ``In free space the quantum vacuum is Lorentz invariant,  so a uniformly moving observer would not see any effect due to motion, but an accelerated observer would.  This is known as the Unruh effect\cite{Unruh:76}.'' Since the vacuum determines the speed of light in the vacuum and the vacuum is at rest with respect to any inertial frame,  it then follows that the speed of light in the vacuum is the same in any inertial frame. The previous statement is one of the two postulates in Einstein's  1905 paper ``On the Electrodynamics of Moving Bodies''\cite{Einstein:05}  in which he introduced special relativity.

In the early universe when the temperature was sufficiently high that it was difficult for charged lepton-antilepton  VFs to bind into charged lepton-antilepton atoms,  the number density of charged lepton-antilepton atoms that are VFs would have  been less than today.  With fewer bound, charged lepton-antilepton VFs per unit volume to slow the progress of photons, the speed of light would have been greater. Moffat\cite{Moffat:93} has pointed out that inflation\cite{Guth:81} with the speed of light much greater in the early universe than it is now allows ``all regions in the universe to be causally connected'', solving the horizon problem.  A sufficiently high superluminary speed also, ``leads to a mechanism of monopole suppression in cosmology and can resolve the flatness problem.''

The fine-structure constant $\alpha$, which is defined by
\begin{equation}\label{eqn:3}
\alpha \equiv \frac {e^2}{4 \pi \epsilon_0\hbar c}\,,
\end{equation}
has historically attracted theoretical interest because it is dimensionless. The fine-structure constant was  introduced by  Sommerfeld in 1916\cite{Sommerfeld:16}  to  quantify relativistic corrections in the Bohr theory of the hydrogen atom and is a measure of the strength of the electromagnetic interaction. However, Sommerfeld was not the first  to recognize the importance of the combined symbols used to define $\alpha$: in the period from 1905-1910\cite{Kragh:03b} both Planck\cite{Hermann:71} and Einstein\cite{Einstein:09} discussed the fact that Planck's constant $h$ and $e^2/(4\pi \epsilon_0 c)$ had the same dimensions.   $\alpha$ grows logarithmically with energy scale, and equals 1/137.036$\dots$ at zero energy scale.   Using the formula for $\epsilon_0$ in   Equation \eqref{eqn:1} and the formula for $c$ in   Equation \eqref{eqn:2},  Equation \eqref{eqn:3} becomes
\begin{equation}\label{eqn:4}
\frac{1}{\alpha} \cong 8^2\sqrt{3\pi/2}= 138.93\dots \, .
\end{equation}
The experimental value for $1/\alpha$ is 1.4\% less than the value calculated in   Equation \eqref{eqn:4}.

This article is organized as follows: In \S2 examples of attempts to calculate the fine-structure constant are briefly discussed. In \S3  properties of charged lepton-antilepton VFs are discussed, including the energy source for their creation and a proof of their existence.
In \S4 the role that the Heisenberg uncertainty principle plays in describing VFs is discussed, yielding an ansatz for the number density of VFs.  \S5  discusses  how  VFs act as harmonic oscillators. In \S6 dielectric properties of the vacuum are calculated and discussed in five subsections:  In \S6.1 the interaction Hamiltonian describing  the interaction of photons with VFs  is constructed, in \S6.2 a quantum calculation of the polarization of VFs is performed, and in \S6.3 a general formula for the permittivity $\epsilon_0$  of the vacuum is derived.  In \S6.4 and \S6.5 the contributions to $\epsilon_0$ from charged lepton-antilepton vacuum fluctuations and  quark-antiquark VFs are, respectively, calculated and discussed. Finally, in \S6.6 the permittivity $\epsilon_0$ of the vacuum is calculated. The process that primarily determines the value of $\epsilon_0$, photon capture by atoms consisting of charged lepton-antilepton vacuum fluctuations and the subsequent decay of these atoms, is discussed.

\section{Attempts to calculate the fine-structure constant}
\label{sec:2}

In 1988 Richard Feynman wrote as follows about the fine-structure constant: ``It has been a mystery ever since it was discovered more than fifty years ago, and all good theoretical physicists put this number up on their wall and worry about it.$\dots$  It's one of the greatest damn mysteries of physics: a magic number that comes to us with no understanding by man. You might say the `hand of God' wrote that number, and `we don't know how He pushed his pencil.' ''\cite{Feynman:88}

Kragh \cite{Kragh:03} writes that there are three reasons that the fine-structure constant is not thought to be as important today as it was in the early decades after its discovery: (a) $\alpha$ is only one of four coupling constants, the other three being the gravitational, weak, and strong. (b)  As mentioned previously, $\alpha$ grows logarithmically with energy scale, and equals 1/137.036$\dots$ only at zero energy scale. (c) There is evidence that the fine-structure constant is not constant over time\cite{Murphy:01}. On the other hand, in Kinoshita's review article\cite{Kinoshita:96} written 80 years after Sommerfeld's introduction of $\alpha$, he points out that $\alpha$ ``characterizes the whole range of physics including elementary particle, atomic, macroscopic, and microscopic systems.''

In the past there have been two general theoretical methods for attempting to calculate the fine-structure constant. (a) physics  nonsense: a physics-based discussion that makes no sense, but nevertheless yields a numerical expression for $\alpha$. (b) numerical coincidence (numerology): a number is found that closely approximates the experimental value for $\alpha$.

In the first category, the most famous physics nonsense determination of $\alpha$ was by Eddington in about 1930.  At that time there were only two known ``elementary'' particles, the electron and the proton. Dirac had incorrectly assumed that a proton is a hole in the, generally filled, sea of negative energy states of the electron. Eddington started with Dirac's theory that uses a $4 \times 4$ matrix to describe an electron. Eddington also used a $4 \times 4$ matrix to describe the proton. To describe both particles requires a $16 \times 16$ matrix, which, if symmetric, would have 136 independent elements. Later, when $1/\alpha$ was found to be closer to 137, Eddington stated that the orbital motion of the electron about the proton added one more degree of freedom, bringing the total to 137.  The physics itself is incorrect, and even if it were correct, there is no reason that the number of degrees of freedom should equal the reciprocal of $\alpha$. A summary of Eddington's approach was given by Birge\cite{Birge:34} in 1934 in which  the  value  Eddington calculates for $\alpha$ is characterized as simply being a result of numerology.

Hans Bethe wrote,``Beck, Riesler and I were at Cambridge on fellowships and had listened to Eddington's unbelievable talk about the number 137''.  To make fun of Eddington's ``derivation'',  they submitted the article \textit{On the Quantum theory of the Temperature of Absolute Zero}\cite{Beck:31}. They had noticed the numerical coincidence that $-273=-(2\times 137-1)$ so they proposed the  following  equation relating the temperature $T_0$ of absolute zero  to  $\alpha$: 
\begin{equation}\label{eqn:5}
T_0=-( \frac{2}{\alpha}-1)\; \mbox{degrees Celsius}\,.
\end{equation}
They then wrote,``Putting $T_0=-273^\circ$, we obtain  for $1/\alpha$ the value 137 in perfect agreement within the limits of accuracy with the value obtained by totally independent methods.''  Of course, the value of absolute zero depends on the temperature scale while $\alpha$ is independent of units. The editor of Die Naturwissenschaften approved the article, and it was published in January, 1931.   In March, 1931, a ``Correction'' was published in Die Naturwissenschaften  stating that the article ``was intended to characterize a certain class of papers in theoretical physics of recent years which are purely speculative and based on spurious numerical agreements.''  

Apparently neither the criticism by Bethe et al. nor Birge had a significant affect on Eddington's self-assurance that he had theoretically calculated  the value of $\alpha$.  Eddington incorporated the above argument into his 1936 book \cite{Eddington:36} and continued to ``clarify'' his method for calculating his value for $\alpha$  in his final book\cite{Eddington:48} published after his death in 1944. In 1994 Kilmister\cite{Kilmister:94} published an analysis of Eddington's work, noting that Eddington always used the term \textit{fine-structure constant} for $1/\alpha$, rather than for $\alpha$.

The second group of calculations rely on a numerical coincidence. For  example, Allen\cite{Allen:15} noted that the mass $m_e$ of the electron divided by 1 atomic mass unit (u) satisfies
\begin{equation}\label{eqn:6}
\frac{m_e}{u}\cong 10\alpha^2\,.
\end{equation}
Other examples consist of   formulas involving integers, $\pi$, square roots, cube roots, etc. that yield a number close to the experimental value for $\alpha$. Such an example by  Wyler\cite{Wyler:69}, which is associated with the symmetric space of the group of Maxwell's  equations, is very close to the accepted value, $1/\alpha= 137.03599\dots$.
\begin{equation}\label{eqn:7}
\frac{1}{\alpha}\cong \frac{16\pi^3}{9}\sqrt[4]{\frac{5!}{\pi}}\cong 137.03608\,.
\end{equation}
So many  examples similar to    Equation \eqref{eqn:7} are now known that it is likely that no physical insight into the value of $\alpha$ is to be gained from them.

\section{Vacuum fluctuations of charged lepton-antilepton pairs}
\label{sec:3}

The appearance of VFs is a stochastic process: as such, either VFs appear on mass shell or they don't appear at all. As discussed previously, a charged lepton-antilepton VF  that results from a fluctuation of the Dirac field will appear in the vacuum as a transient atom.   During the time while such an atom exists, it can interact with a photon. The fundamental interaction of a photon with a VF is the capture of the photon by  the charged lepton-antilepton VF. 

Field theory provides a simple explanation for the source of the energy available for the creation of a VF and a proof that VFs must exist.  The structure of an atom that is a VF does not play a significant role in the discussion of the energy required to produce the VF.  (Of course, the structure of the VF, which results from the electromagnetic interaction of the charged lepton-antilepton pair, is important for  the calculation of the decay rate of the atom that is a VF.)  Thus an atom consisting of a charged lepton and antilepton in its  ground state with zero angular momentum can, as far as its creation is concerned, be approximately represented  by a free, neutral,  spin-0, Klein-Gordon field $\phi(x)$, as first suggested by Pauli and Weisskopf\cite{Pauli:94} and elaborated on by Wentzel\cite{Wentzel:03}. Using a field to describe a particle with internal degrees of freedom is discussed in Ref.\cite{Thirring:58}; representing a particle by a quantum field is essential to understanding the mathematical structure of a VF of a quantum field. Indeed, it is the quanta of a free field that behave as free particles\cite{Roman:69}.

The energy available for the creation of VFs is the energy in the vacuum, called the zero-point energy, and is given by the vacuum expectation value $\langle 0|H|0\rangle$ of the Hamiltonian $H$ for a free, neutral  Klein-Gordon field\cite{Peskin:95}. If there is no cutoff in momentum, the vacuum expectation $\langle 0|H|0\rangle$ is infinite; if there is a cutoff, the vacuum energy is finite, but still large. As Peskin and Schroeder point out, since experiments measure only energy differences, the energy associated with VFs cannot be measured directly in elementary particle experiments\cite{Peskin:95}.

The {\it presence} of Klein-Gordon VFs, however, is easily established.  To show that vacuum fluctuations of charged lepton-antilepton pairs must exist, first note that for a  free field $\phi(x)$, the vacuum expectation value $\langle 0|\phi(x)|0\rangle=0$. In contrast, the expectation value  of the product of the free field at two different locations $x$ and $x^\prime$ is\cite{Thirring:58,Bjorken:65} $\langle 0|\phi(x)\phi(x^\prime)|0\rangle\neq 0$.  Accordingly, the vacuum expectation of the square of the field $\phi(x)$ deviates from the square of the vacuum expectation value of the field $\phi(x)$. This demonstrates that the free field $\phi(x)$ in the vacuum is nonzero and means that field theory predicts that fluctuations of a free (noninteracting) field occur in the vacuum.

The purpose of the present paper is to demonstrate a measurable effect of the presence of charged lepton-antilepton VFs, namely, the importance of these VFs in establishing the value of $\epsilon_0$.  This calculation does not depend on an absolute value for the zero-point energy.  As Peskin and Schroeder point out, a potential method for establishing (measuring) an absolute value for the zero-point energy in the vacuum lies in understanding the coupling of that energy to gravity via the cosmological constant in EinsteinÕs  equations.  Such a direct comparison seems at present unlikely, since the calculation and measurement appear to differ by a factor of around $10^{120}$\cite{Peskin:95}.

\section{Heisenberg uncertainty principle and  charged lepton-antilepton vacuum fluctuations}
\label{sec:4}

There are three types of charged lepton-antilepton vacuum fluctuations: (a) electron-positron (b) muon-antimuon (c) tau-antitau. The electron-positron VF in the lowest energy level that has zero angular momentum is  parapositronium,   which is a singlet spin state\cite{Deutsch:52,DeBenedetti:54,Hughes:57}.  Initially attention will be restricted to parapositronium since the corresponding calculations for  muon-antimuon and tau-antitau VFs can be obtained by replacing the mass of the electron with the mass of  the muon or tau, respectively.

For parapositronium that is a VF, the Heisenberg uncertainty principle is 
\begin{equation}\label{eqn:16}
\Delta E_{\rm p-Ps}\, \Delta t_{\rm p-Ps}\geq\frac{\hbar} {2} \, .
\end{equation}
Denoting the mass of an electron (or positron) by $m_e$,  $\Delta E_{\rm p-Ps}$ is the energy $2m_ec^2$ for the production  of parapositronium that is a VF \footnote{The binding energy of parapositronium, which is small in comparison with  $2m_ec^2$, is being neglected.}. The minimum uncertainty in time is the average time $\Delta t_{\rm p-Ps}$ that a parapositronium VF exists.  Then  Equation \eqref{eqn:16} yields 
\begin{equation}\label{eqn:17}
\Delta t_{\rm p-Ps}= \frac{\hbar}{4m_ec^2} \, .
\end{equation}
During the time $\Delta t_{\rm p-Ps}$, a beam of light  travels a distance $L_{\rm p-Ps}$ given by
\begin{equation}\label{eqn:18}
L_{\rm p-Ps}= c \,\Delta t_{\rm p-Ps}= \frac{\hbar}{4m_ec} \, .
\end{equation}

Since a parapositronium VF appears from the vacuum at essentially a single location and since nothing can travel faster than the speed of light, while they exist the maximum displacement between the electron and positron in parapositronium  is  $L_{\rm p-Ps}$: for a time  $\Delta t_{\rm p-Ps}/2$ the electron and positron can move apart, and for a time $\Delta t_{\rm p-Ps}/2$ they must move back toward each other in order to annihilate on average at the time $\Delta t_{\rm p-Ps}$.  Since an energy $2m_ec^2$ has already been borrowed from the volume $L_{\rm p-Ps}^3$ of the vacuum for the creation of a parapositronium atom,  it is reasonable to assume that  another parapositronium VF is unlikely to form  in the same volume. Thus for the model being discussed,
\begin{equation}\label{eqn:19}
\mbox{number of parapositronium atoms/volume}  = \frac{1}{L_{\rm p-Ps}^3}\,.
\end{equation}
The result  Equation \eqref{eqn:19} can  immediately be generalized to other charged lepton-antilepton VFs and quark-antiquark VFs.   The number density of charged lepton-antilepton VFs ranges from $1.12 \times 10^{39}$/m$^3$ for electron-positron VFs to   $4.70 \times 10^{49}$/m$^3$ for tau-anti-tau VFs.   In a 6,000 W, CO$_2$ cutting laser with a beam diameter of 0.32 mm, the number density of photons is on the order of $10^{22}$ photons/m$^3$. Even in such an intense laser beam, the number density of charged lepton-antilepton VFs is much  greater than the number density of photons. Thus if a charged lepton-antilepton VF interacts with a photon at all, it essentially always interacts with only one photon.

The number density of atoms or molecules of an ideal gas at STP is $2.68 \times 10^{25}$/m$^3$. It is possible for the number density of VFs to be many orders of magnitude greater than the number density of atoms or molecules in a gas because a VF cannot exert a force. Consider first the electromagnetic force: if a VF has not already absorbed radiation, it cannot spontaneously emit radiation. If it did, the radiated photons would exist after the VF has disappeared back into the vacuum, permanently violating conservation of energy.  If a VF has interacted with a photon, when the VF vanishes back into the vacuum the VF must emit a photon identical to the incident photon in order to conserve energy, momentum, and angular momentum. Since a VF cannot ``permanently'' exchange a photon with either a VF or a physical quantum, it cannot exert a force on either.   Similar arguments establish that VFs cannot exert a force of any type. 

In addition to absorbing and emitting photons, VFs can also interact through the annihilation of a physical particle and a VF or  a physical antiparticle and a  VF.  For example, a physical electron can annihilate with a positron that is part of a VF.  The electron that was also part of the VF then becomes a physical electron with a location different from that of the original, physical electron,  giving rise to zitterbewegung\cite{Thirring:58}.  

\section{Charged lepton-antilepton vacuum fluctuations as harmonic oscillators}
\label{sec:5}
% Section numbering is automatic.

Here attention is again restricted to parapositronium.  In the center-of-mass rest frame,  the relative position of the positron and electron are given by $\vec{r}=\vec{r}_+-\vec{r}_-$, where $\vec{r}_+$ and $\vec{r}_-$ are, respectively, the positions of the positron and electron. Note that $\vec{r}$ points in the direction of the electric dipole moment  of parapositronium.  For the hydrogen atom $\vec{r}$  usually points from the positive nucleus to the electron. However, the Schr\"odinger  equation for parapositronium, which in the center-of-mass  reference frame has a spherically symmetric wave function, is identical to the Schr\"odinger  equation describing states of the hydrogen atom with spherically symmetric wave functions except that the reduced mass $\mu$ is different: for hydrogen the reduced mass is approximately $m_e$, where $m_e$ is the mass of an electron. For  parapositronium the reduced mass is $m_e/2$.  The  non-relativistic  binding energy for parapositronium, $E_{p-Ps}$,  is  obtained from the $n = 1$ binding energy of hydrogen\cite{Schiff:55} just by changing the reduced mass:
\begin{equation}\label{eqn:30}
E_{p-Ps} = -\frac{ (m_e/2)e^4 }{2(4\pi \epsilon_0)^2 \hbar^2} =-\frac{m_e\alpha^2c^2}{4} \,.  \\
\end{equation}

From both the classical\cite{Kramers:24,Rossi:59,Feynman:63} and quantum\cite{Sokolov:84} calculations of the permittivity of  physical matter consisting of atoms (or molecules),  it follows that for the matter to possess permittivity,  the atoms must be able to oscillate when interacting with an electric field (or photons).  Taking the electric field to point in the $x$-direction, since the atom oscillates along the direction of the electric field, only the oscillatory properties of the atom  in the $x$-direction are of significance in the interaction.  Thus, when calculating the permittivity, the interaction of the electromagnetic field with the atom can be described by a one-dimensional potential. 

Similarly to many other systems in physics that can be described by a one-dimensional potential $U(x)$, the parapositronium atom can be expected to oscillate if the potential has a minimum at $x_e$ and can be expanded in a Taylor series about that minimum: 
\begin{subequations}\label{eqn:31}
\begin{equation} \label{eqn:31a}
U(x)=U(x_e ) + \frac{\mathrm{d}U(x)}{\mathrm{d}x} 
\biggl|_{x=x_e} \hspace{-0.7 cm}(x-x_e) +\frac{1}{2!} \frac{{\mathrm{d}}^2 U(x)}{\mathrm{d}x^2}
\biggr|_{x=x_e} \hspace{-0.7 cm}(x-x_e )^2 + \dots. 
\end{equation}
In the above formula $x = x_+  -  x_-$.  At $x_e$ there is a relative minimum of the potential so $\frac{\mathrm{d}U(x)}{\mathrm{d}x}$ is zero at $x_e$. Choosing  the origin of the $x$-axis at the equilibrium position $x_e$ so that $x_e=0$,  Equation \eqref{eqn:31a} can be rewritten as 
\begin{equation} \label{eqn:31b}
U(x)\cong U(x_e=0 ) + \frac{1}{2}Kx^2\,.
\end{equation}
\end{subequations}
Equation \eqref{eqn:31b} is the  equation for a one-dimensional harmonic oscillator potential with a spring constant $K=\frac{\mathrm{d}^2U(x)}{\mathrm{d}x^2}\big|_{x=x_e}$. The constant term $U(x_e=0)$ just shifts all energy levels by the same amount.

As pointed out by Feynman\cite{Feynman:64}, when interacting  with an electric field,  an atom in its ground state interacts with the electric field as if it were a harmonic oscillator with the first two  energy levels separated by the binding energy of the atom. Since adjacent harmonic oscillator energy levels are separated by an energy $\hbar \omega^0$, from the expression   Equation \eqref{eqn:30} for the binding energy $E_{p-Ps}$ of parapositronium,
\begin{equation}\label{eqn:32}
\omega^0=\frac{|E_{p-Ps}|}{\hbar}=\frac{m_e\alpha^2c^2}{4\hbar} \,.
\end{equation}
The spring constant $K$ of the harmonic oscillator corresponding to parapositronium is  
\begin{equation}\label{eqn:33}
K=\mu (\omega^0)^2=\frac{m_e}{2}\left(\frac{m_e\alpha^2c^2}{4\hbar}\right)^2 \,.
\end{equation}
Using the formula for the energy\cite{Schiff:55}  of a harmonic oscillator in one dimension, $E=\hbar \omega^0(n+1/2)\,, \;n=0,1,2\dots $, the energy eigenvalues $E_n^0$ of parapositronium are  
\begin{equation}\label{eqn:34}
E_n^0=U(x_e=0)+\hbar \omega^0(n+\frac{1}{2})\,.
\end{equation}
The harmonic oscillator matrix element that appears in the calculation of the permittivity does not depend on the constant term $U(x_e=0)$  in the potential, so there is no need to determine a specific value.

In the center-of-mass rest frame of the parapositronium VF, the  parapositronium VF is described by the harmonic oscillator Hamiltonian $H^0$,
\begin{equation}\label{eqn:35}
H^0=\frac{1}{2}\mu\left(\frac{{\rm dx}}{{\rm d}t}\right)^2+U(x_e=0)+ \frac{1}{2}\mu (\omega^0)^2 x^2\,,
\end{equation}
where $\omega^0$ is defined in   Equation \eqref{eqn:32}. 

\section{Calculation of the permittivity of the vacuum}
\label{sec:6} 

\subsection{Interaction Hamiltonian for photons interacting with vacuum fluctuations}
\label{subsec:6.1} 

The one-dimensional harmonic oscillator described by the Hamiltonian  Equation \eqref{eqn:35} can undergo an electric dipole transition between its ground state and first excited state because the initial and final states have opposite parity. The interaction Hamiltonian  describing the dipole interaction of a photon  with a VF follows:

The dipole moment (operator) $p_x$ of an atom in the presence an electromagnetic wave with its electric field in the $x$-direction is  $p_x=e(x_+-x_- )=ex$, where $e$ is the magnitude of the charge on an an electron.  The Hamiltonian $H^1$ describing the interaction of the electric dipole of the atom with the electromagnetic wave is
\begin{equation}\label{eqn:202}
H^1=-\mathbf{p}\cdot\mathbf{E(t)}=-exE_0\cos \omega t\,.
\end{equation}

The two Hamiltonian  equations for the Hamiltonian $H=H^0+H^1$ can be  combined to yield
\begin{equation}\label{eqn:203}
-Kx+eE_0\cos \omega t =\mu \frac{{\rm d}^2x(t)}{{\rm d}t^2}\,,
\end{equation}
which is just Newton's second law. The classical calculation of the permittivity of ordinary matter is based on  Equation \eqref{eqn:203}  with one exception: for ordinary matter there is also a phenomenological term proportional to velocity that describes damping that results from radiation and collisions. For gases damping is often very small and can be neglected.  While neglecting damping  is an approximation for physical particles in a  dielectric, it is exactly true for VFs. VFs can neither radiate energy nor lose energy in collisions with other quanta. If they did, after they vanished they would permanently leave behind energy,  violating the principle of conservation of energy.

To describe the interaction of VFs with an electric field, one change has to be made to $H^1$.  If an atom is continually interacting with photons from an electric field $E_0\cos \omega t\;\mathbf{\hat{x}}$, then the interaction $H^1$ of the atom with the field is given by  Equation \eqref{eqn:202}. As discussed in \S \ref{sec:4}, even in an intense laser beam the number density of photons is much less than the number density of VFs. Thus the probability that at a given time  a VF interacts with more than one photon is very small.  That is, if a VF interacts with an electromagnetic wave, it almost always does so with only one photon. 

If a VF  absorbs a photon at time $t_i$,  the photon vanishes at that instant. The electric field at the moment of interaction is $E(t_i) = E_0\cos \omega t_i \equiv \mathbb{E}_0$, implying that the VF interacts with the electric field $\mathbb{E}_0\,\mathbf{\hat{x}}$. Thus, for VFs  Equation \eqref{eqn:202} becomes
\begin{equation}\label{eqn:204}
H^{1 ^{V\!F}}=-exE_0\cos \omega t_i \equiv -ex\mathbb{E}_0\,.
\end{equation}
As a part of the capture process, a  parapositronium  atom will, to some extent, be excited by $\mathbb{E}_0$ to resonate at its characteristic frequency $\omega^0$, as discussed in the following section.

\subsection{Quantum calculation of the polarization of vacuum fluctuations}
\label{subsec:6.2} 

Because photons  interact with atoms that are VFs somewhat similarly to the  the way that they interact with  ordinary atoms, it is possible to calculate the permittivity $\epsilon_0$  of the vacuum  using techniques  similar to those employed for calculating the permittivity $\epsilon$ of a physical dielectric.  In the vacuum the interactions that primarily contribute  to the value of $\epsilon_0$ are photon capture by charged lepton-antilepton VFs bound into the lowest energy state that has zero angular momentum. The charged lepton-antilepton pair quickly annihilate, emitting a photon identical to the incident, captured photon.

Let $\psi_n^0(x,t)$ be solutions to the unperturbed Schr\"odinger  equation,
\begin{equation}\label{eqn:205}
i\hbar\frac{\partial \psi_n^0(x,t)}{\partial t}=H^0\psi_n^0(x,t)=E_n^0\psi_n^0(x,t)\,,
\end{equation}
where   $E_n^0$ and  $H^0$ are given in  Equation \eqref{eqn:34} and  Equation \eqref{eqn:35}, respectively.  From    Equation \eqref{eqn:205} it follows that the energy dependence of $\psi_n^0(x,t)$ can be factored,
\begin{equation}\label{eqn:2051}
 \psi_n^0(x,t)=e^{-E_n^0 t/\hbar}\psi_n^0(x)\,.
\end{equation}
The ``exact'' wave function  $\psi(x,t)$ satisfies the Schr\"odinger  equation, 
\begin{equation}\label{eqn:206}
i\hbar\frac{\partial \psi(x,t)}{\partial t}=(H^0+H^{1^{V\!F}})\,\psi(x,t)\,.
\end{equation}

Perturbation theory is now used to calculate $\psi(x,t)$ to first order in the perturbation $H^{1^{V\!F}}\hspace{-0.1 cm}$:  the ``exact'' wave function  $\psi(x,t)$ is written as 
\begin{align}\label{eqn:207}
\psi(x,t)&=a_0(t)\psi_0^0(x,t)+a_1(t)\psi_1^0(x,t)\,,\nonumber\\
&=a_0(t)e^{-iE_0^0 t/\hbar}\psi_0^0(x)+a_1(t)e^{-iE_1^0 t/\hbar}\psi_1^0(x)\,.
\end{align}
Substituting  Equation \eqref{eqn:207} into  Equation \eqref{eqn:206},
\begin{align}\label{eqn:208}
i\hbar\frac{\rm{d}a_0(t)} {\rm{d}t}e^{-iE_0^0 t/\hbar}\psi_0^0(x)&+i\hbar\frac{\rm{d}a_1(t)} {\rm{d}t}e^{-iE_1^0 t/\hbar }
\psi_1^0(x)\\ \nonumber
&=H^{1^{V\!F}}a_0(t)e^{-iE_0^0 t/\hbar}\psi_0^0(x)+H^{1^{V\!F}}a_1(t)e^{-iE_1^0 t/\hbar}\psi_1^0(x)\,.
\end{align}
 Using the orthogonality relations,
\begin{equation}\label{eqn:2081}
\int_{-\infty}^\infty{\rm d}x\,\psi_{n^\prime}^{0*}(x)\,\psi_n^0(x)=\delta_{n^\prime,n}\,,
\end{equation}
and noting that $\psi_0^{0}(x)$ and $\psi_1^{0}(x)$ have opposite parity  while $H^{1 ^{V\!F}}\equiv -ex\mathbb{E}_0$ has negative parity,  equations for $\frac{\rm{d}a_0(t)} {\rm{d}t}$ and $\frac{\rm{d}a_1(t)} {\rm{d}t}$ are immediately obtained from  Equation \eqref{eqn:208}:
\begin{subequations} \label{eqn:2082}
\begin{align}
 \label{eqn:2082a}
\frac{\rm{d}a_0(t)} {\rm{d}t}&=a_1(t) \frac{ie\mathbb{E}_0}{\hbar}\langle x\rangle_{0,1}e^{-i\omega^0t}\,, \\
 \label{eqn:2082b}
\frac{\rm{d}a_1(t)} {\rm{d}t}&=a_0(t) \frac{ie\mathbb{E}_0}{\hbar}\langle x\rangle_{1,0}e^{i\omega^0t}\,. 
\end{align}
\end{subequations}
In the above  equation
\begin{equation}\label{eqn:2083}
\langle x\rangle_{n^\prime,n}\equiv \int_{-\infty}^\infty{\rm d}x\,\psi_{n^\prime}^{0*}(x)\,x\,\psi_n^0(x)
\end{equation}
and, as discussed in \S5,
\begin{equation}\label{eqn:2084}
\omega^0\equiv\frac{1}{\hbar}(E_1^0-E_0^0)\,.
\end{equation}

Letting the time $t_i$ that the photon interacts with the parapositronium VF  be infinitesimally greater than $t=0$ implies that
\begin{equation}\label{eqn:2085}
a_0(t=0)=1\;\; \mbox{and}\;\;  a_1(t=0)=0 \,,
\end{equation}
because the parapositronium VF is initially in its ground state, which, when interacting with a photon, is represented by the ground state $\psi_0^0(x,t)$ of the harmonic oscillator. Using  Equation \eqref{eqn:2085},  Equation \eqref{eqn:2082b} becomes
\begin{equation}\label{eqn:2086}
\frac{\rm{d}a_1(t)} {\rm{d}t}= \frac{ie\mathbb{E}_0}{\hbar}\langle x\rangle_{1,0}e^{i\omega^0t}\,,
\end{equation}
which can immediately be integrated to yield
\begin{equation}\label{eqn:2087}
a_1(t)= \frac{e\mathbb{E}_0}{\hbar \omega^0}\langle x\rangle_{1,0}e^{i\omega^0t}\,.
\end{equation}
Note that   Equation \eqref{eqn:2087} satisfies   Equation \eqref{eqn:2085} because the interaction has not yet occurred at time $t_i=0$.  If  the formula for $a_1(t)$ in   Equation \eqref{eqn:2087} is substituted into  Equation \eqref{eqn:2082a}, a correction to $a_0(t)$ is obtained that is on the order of $(e\mathbb{E}_0)^2$.  Thus from  Equation \eqref{eqn:2085} and  Equation \eqref{eqn:2087}, to order $e\mathbb{E}_0$ the ``exact'' wave function $\psi(x,t)$ is given by
\begin{align}\label{eqn:2088}
\psi(x,t)&=e^{-iE_0^0 t/\hbar}\psi_0^0(x)+ \frac{e\mathbb{E}_0}{\hbar \omega^0}\langle x\rangle_{1,0}e^{i\omega^0t}e^{-iE_1^0 t/\hbar}\psi_1^0(x)\,,\nonumber\\
&=e^{-iE_0^0 t/\hbar}[\psi_0^0(x)+ \frac{e\mathbb{E}_0}{\hbar \omega^0}\langle x\rangle_{1,0}\psi_1^0(x)]\,.
\end{align}

From   Equation \eqref{eqn:202} and   Equation \eqref{eqn:204} the expectation value $\langle p^{V\!F}\!\rangle$ of the electric dipole moment in the state characterized by $\psi(x,t)$ is
\begin{align}\label{eqn:218}
&\langle p^{V\!F}\!\rangle=\int_{-\infty}^\infty{\rm d}x\,\psi^*(x,t)\,(ex)\,\psi(x,t)\,, \nonumber \\
&=\int_{-\infty}^\infty{\rm d}x \,e^{iE_0^0 t/\hbar}[\psi_0^{0*}(x)+ \frac{e\mathbb{E}_0}{\hbar \omega^0}\langle x\rangle_{1,0}\psi_1^{0*}(x)](ex)
e^{-iE_0^0 t/\hbar}[\psi_0^0(x)+ \frac{e\mathbb{E}_0}{\hbar \omega^0}\langle x\rangle_{1,0}\psi_1^0(x)]\,,\nonumber\\
&\cong \int_{-\infty}^\infty{\rm d}x\, [\psi_0^{0*}(x)(ex) \frac{e\mathbb{E}_0}{\hbar \omega^0}\langle x\rangle_{1,0}\psi_1^0(x)+
 \frac{e\mathbb{E}_0}{\hbar \omega^0}\langle x\rangle_{1,0}\psi_1^{0*}(x)(ex)\psi_0^0(x)]\,, \nonumber\\
 &= \frac{2e^2\mathbb{E}_0}{\hbar\omega^0}\langle x\rangle_{1,0}^2\,.
\end{align}
To obtain the final equality, phases were chosen so that $\psi_0^0(x)$ and $\psi_1^{0}(x)$ are both real. Using the one-dimensional harmonic oscillator result, 
\begin{equation}\label{eqn:221}
\langle x\rangle_{1,0}= \sqrt{\frac{\hbar}{2\mu\omega^0}} \,,
\end{equation}
it immediately follows from   Equation \eqref{eqn:218} that
\begin{equation}\label{eqn:222}
\langle p^{V\!F}\!\rangle_{0,0}=\frac{(e^2/\mu)\mathbb{E}_0}{ (\omega^0)^2} \,.
\end{equation}
Describing an oscillator as a harmonic oscillator, the quantum formula  for the expectation value of the electric dipole of a VF agrees with the classical formula for the electric dipole of an atom of ordinary matter with two exceptions that have already been discussed:  (1) Since VFs can neither radiate energy nor loose energy in a collision, there is no damping term.  (2)  Because VFs essentially always  only interact with a single photon,  only the value of the electric field  at the instant of interaction is relevant so there is no dependence on  the  angular frequency $\omega$ of the incident photon.

To allow for the possibility that there is more than one type of atom made from  particle-antiparticle VFs, an index $j$ is added:  accordingly,  the charge $e \rightarrow q_j$, the reduced mass $\mu \rightarrow \mu_j$, and the resonant frequency $\omega^0 \rightarrow \omega_j^0$.  
\begin{equation}\label{eqn:223}
\langle p^{V\!F}_j\!\rangle_{0,0}=\frac{(q_j^2/\mu_j)\mathbb{E}_0}{(\omega_j^0)^2} \, .
\end{equation}

\subsection{General formula for the permittivity $\epsilon_0$ of the vacuum}
\label{subsec:6.3}

In a dielectric\cite{Jackson:99} the electric displacement $D(t)$ satisfies 
\begin{equation}\label{eqn:224}
D(t)=\epsilon E(t)=\epsilon_0 E(t) +P(t)\,,
\end{equation}
where $\epsilon$ is the permittivity of the dielectric, $E(t)$ is the electric field, and $P(t)$ is the electric polarization (dipole moment per unit volume).  $P(t)$  can be expressed in terms of the individual dipole moments $p_j(t)$:
\begin{equation}\label{eqn:225}
P(t)=\sum_j N_j p_j(t)\,.
\end{equation}
In   Equation \eqref{eqn:225} $N_j$ is the number of oscillators per unit volume of the $j^{\rm th}$ variety that  are available to interact.  In a uniform, classical electric field, the field is everywhere $E_0\cos \omega t$.  Since the electric  polarization is proportional to  the electric field, the electric field cancels out of  Equation \eqref{eqn:225}.

The electric polarization $P(t)$ is responsible for the increase from  $\epsilon_0 E(t)$ to $\epsilon E(t)$  because of photons interacting  with oscillators in the material dielectric and results entirely from polarization of the atoms, molecules or both in  the dielectric. It then follows that in  the vacuum $\epsilon_0 E(t)$ must result entirely from the electric polarization  $P^{V\!F}(t)$ of atoms, molecules, or both that are VFs.  Thus, 
\begin{subequations} \label{eqn:226}
\begin{equation}\label{eqn:226a}
\epsilon_0 E(t)  = P^{V\!F}(t)\,, 
\end{equation}
or
\begin{equation}\label{eqn:226b}
\epsilon_0  = \frac{P^{V\!F}(t)}{E(t) }\,.
\end{equation}
\end{subequations}
The measurement of the permittivity $\epsilon_0$ of the vacuum occurs over a time interval $\Delta t$. For any time $t_i$ in the interval $\Delta t$ for which a photon-VF interaction occurs, the electric field at that instant is $E(t_i) = E_0\cos \omega t_i \equiv \mathbb{E}_0$.  As shown in  Equation \eqref{eqn:223}, the polarization density is proportional to $\mathbb{E}_0$, implying that the instantaneous value of the electric field will cancel out in  Equation \eqref{eqn:226}.  From  Equation \eqref{eqn:226b},  Equation \eqref{eqn:225} and  Equation \eqref{eqn:223}
\begin{equation}\label{eqn:227}
\epsilon_0=\sum_j N^{V\!F}_j\frac{ p_j^{V\!F}(t_i)}{\mathbb{E}_0}=\sum_j N^{V\!F}_j \frac{\langle p_j^{V\!F}\!\rangle_{0,0}}{\mathbb{E}_0}=\sum_j N^{V\!F}_j \frac{(q_j^2/\mu_j)}{(\omega_j^0)^2}\,.
\end{equation}
Since the value of the electric field has canceled out of  Equation \eqref{eqn:227}, $\epsilon_0$ is independent of the polarizing electric field, including its frequency.  As a result, there is no dispersion in the vacuum.

Contributions from various types of VFs  must now be summed over in   Equation \eqref{eqn:227} to obtain  the formula for $\epsilon_0$.  The three types of VFs  first considered are atomic, bound states of a charged lepton and antilepton, namely, parapositronium, muon-antimuon bound states, and tau-antitau bound states. Quark-antiquark states will  be discussed later.   Initially attention is restricted to a parapositronium  VF.

\subsection{Contribution to $\epsilon_0$ from charged lepton-antilepton vacuum fluctuations}
\label{subsec:6.4} 

The progress of a photon traveling through the vacuum is slowed when it interacts with and has a polarizing effect on a VF consisting of a charged lepton and antilepton bound into an atom in its ground state.  The fundamental interaction of a photon with a VF is the capture of the photon by  the charged lepton-antilepton VF.  Labeling the initial (incident) and final (emitted) photons, respectively, by $\gamma_i$ and $\gamma_f$, to lowest order the two Feynman diagrams that contribute to the process $\gamma_i$+charged lepton-antilepton VF $\rightarrow \gamma_f$ are shown in Fig.~\ref{fig:1}.
\setcounter{figure}{0}
\begin{figure}[h]
\vspace{-0.6 cm}
\begin{tabular}{cc}
\includegraphics[width=88mm]{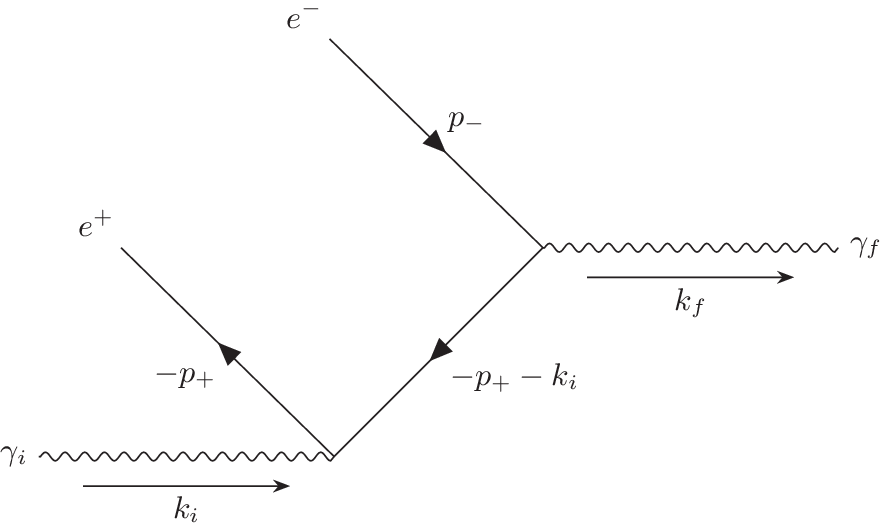}\hspace{0.5 cm}&\includegraphics[width=63mm]{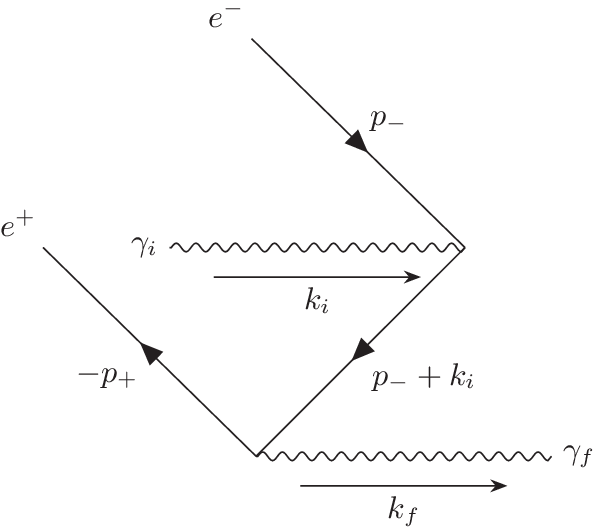}\\
(a)&(b)\\
%&\vspace{-0.5 cm}(a)&\vspace{-0.5 cm}(b)\\
\end{tabular}
%\vspace{1.0 cm}
\caption{ (a) Photon $\gamma_i$ interacts with an antilepton that then annihilates with an lepton, emitting  photon $\gamma_f$. (b) Photon  $\gamma_i$ interacts with a lepton that then annihilates with an antilepton, emitting  photon $\gamma_f$.}
\label{fig:1}
\end{figure}
In the  diagrams $p_-, p_+, k_i,$ and $k_f$ are, respectively, the four-momenta of the lepton, antilepton, initial  photon, and final photon.  

A quantum electrodynamics calculation\cite{Mainland:18a}  determines the decay  rate $\Gamma$ of this quasi-stationary atom.   Using $\Gamma$, the effect of this interaction on the permittivity of the vacuum can be calculated. Coherence between the initial and final state is maintained.  As required by conservation of energy and momentum, when the photon-excited, quasi-stationary state annihilates, the energy and momentum that were originally borrowed from the vacuum are returned to the vacuum.  An isolated, ordinary atom consisting of a charged lepton-antilepton pair is kinematically forbidden from capturing a photon,  annihilating, and then emitting a single photon; however, because the kinematics is different for the capture and release of a photon by a VF, the process is kinematically allowed for  charged lepton-antilepton VFs.\
 
The electromagnetic decay rate $\Gamma_{\rm p-Ps}$  for a  parapositronium VF after it has interacted with the incident photon to form a  quasi-stationary  state is\cite{Mainland:18a}
\begin{equation}\label{eqn:228}
\Gamma_{\rm p-Ps} = \frac{\alpha^5 m_e c^2}{ \hbar} \, .
\end{equation}
The above rate is twice the decay rate of ordinary parapositronium into two photons\cite{Wheeler:46,Jauch:55}.

The probability that an excited  parapositronium VF has not decayed during a time $t$ is  $e^{-\Gamma_{\rm p-Ps}t}$, and the probability that it has  decayed electromagnetically  is $1-e^{-\Gamma_{\rm p-Ps}t}$.  The quantity $N_j^{VF}$ in   Equation \eqref{eqn:227} for a parapositronium VF is the number of  parapositronium VFs  per unit volume  with which a photon actually interacts.  At equilibrium the average rate at which a parapositronium VF  absorbs a photon equals the average rate for a  parapositronium VF to  annihilate and emit a photon. As a  consequence, the average probability that  parapositronium absorbs a photon during a time  $t$  is $1-e^{-\Gamma_{\rm p-Ps}t}$.

 For a parapositronium VF  the quantity $N^{VF}_j$, denoted  by $N_{\rm p-Ps}$,  is the number density of   parapositronium VFs  multiplied by the probability that a  parapositronium VF  will absorb an incoming photon during the lifetime $\Delta t_{\rm p-Ps}$ of the parapositronium VF: 
\begin{align}\label{eqn:229}
N_{\rm p-Ps}^{V\!F}&\cong \frac{1}{L_{\rm p-Ps}^3}\times(1- e^{-\Gamma_{\rm p-Ps}\, \Delta t_{\rm p-Ps}}) \,.
\end{align}
Since $\Gamma_{\rm p-Ps}\,\Delta t_{\rm p-Ps} \ll1$,  the term $1- e^{-\Gamma_{\rm p-Ps}\, \Delta t_{\rm p-Ps}}$ is very nearly equal to $\Gamma_{\rm p-Ps}\,\Delta t_{\rm p-Ps}$. Thus 
\begin{align}\label{eqn:230}
N_{\rm p-Ps}^{V\!F} \cong \frac{1}{L_{\rm p-Ps}^3}\times \Gamma_{\rm p-Ps}\Delta t_{\rm p-Ps}=  \frac{\alpha^5}{4}{\left ({\frac{4 m_e c}{\hbar}} \right )}^3\,.
\end{align}

To better understand  Equation \eqref{eqn:230}, for a particular atom (or molecule) let $N$ be the number of atoms per unit volume, let $\Gamma$ be the decay rate of a photon-excited atom  into the atom  in its ground state plus a photon, and let $t$ be the average time that the atom exists.  For ordinary matter $N_j$ in   Equation \eqref{eqn:225} is $N$, but for a vacuum fluctuation it is $N\Gamma t$ as given in  Equation \eqref{eqn:230}.  How does this difference arise?  Making no assumptions about the magnitude of $\Gamma t$ and using the  logic that led to  Equation \eqref{eqn:229}, $N_j=N(1-e^{-\Gamma t})$.  If the atom is stable, the lifetime $t$ is infinite so $\Gamma t$ is infinite, and $N_j=N$ as expected. On the other hand, if the lifetime is sufficiently small   that  $\Gamma t \ll1$, as is the case for a parapostronium VF, $N^{VF}_j\cong N\Gamma t$ as given in  Equation \eqref{eqn:230} \footnote{As pointed out by A. Bohm\cite{Bohm:81}, a physical state prepared in a scattering experiment contains background information about the reaction in which it was created.  The above description of a single photon interacting with a VF clearly depicts such a circumstance despite the fact that the details of the discussion involve a decay rate $\Gamma$. Usually, a decay rate such as $\Gamma_{p-Ps}$ is associated only with a stochastic process.  In each individual interaction of a photon with a VF, the coherence of the final photon with the initial photon is assured, as promised at the beginning of \S\ref{subsec:6.4}.}. 

Substituting  Equation \eqref{eqn:32} and  Equation \eqref{eqn:230} into  Equation \eqref{eqn:227},
\begin{equation}\label{eqn:231}
\epsilon_0 \cong  \sum_j  \frac{ 8^3\alpha e^2}{\hbar c} \, .
\end{equation}
Note that the mass of the electron has cancelled from the expression for $\epsilon_0$, implying that bound muon-antimuon  and tau-antitau VFs each contribute the same amount to the value of  $\epsilon_0$ as parapositronium VFs contribute. Including the  contributions from the three types of charged, bound lepton-antilepton VFs yields
\begin{equation}\label{eqn:232}
\epsilon_0 \cong 3\,  \frac{ 8^3\alpha e^2}{\hbar c}\, +\left(\substack{\mbox{any contribution from}\\ \\\mbox{quark-antiquark VFs.}}\right)
\end{equation}

\subsection{Contribution to $\epsilon_0$   from quark-antiquark vacuum fluctuations}
\label{subsec:6.5} 

The only other charged particle-antiparticle VFs that might exhibit elastic behavior are quark-antiquark VFs; however,  as will now be shown, contributions to $\epsilon_0$   from  quark-antiquark  VFs  is substantially reduced in comparison with those from charged lepton-antilepton VFs.  

First consider the heavy quarks  $Q=c, b,\, {\rm or}\; t$, where it is appropriate to think in terms of static quark potentials for $Q{\bar Q}$ bound states. Let $m_{Q{\bar Q}}$ be the mass  of the least massive $Q{\bar Q}$ bound state that has zero angular momentum. From the Heisenberg uncertainty principle,  the  $Q{\bar Q}$ bound state will have an average lifetime $\Delta t_{Q{\bar Q}}=\hbar/(2M_{Q{\bar Q}}c^2)$. During the time $\Delta t_{Q{\bar Q}}$ light will travel a distance $L_{Q{\bar Q}}=c \Delta t_{Q{\bar Q}}$.   As previously noted, the decay rate of a photon-excited parapositronium VF into a photon is twice the decay rate of ordinary parapositronium into two photons. Since the decay rate of  a photon-excited $Q{\bar Q}$ VF  into a photon is not known, it is approximated by  twice the decay  rate $\Gamma_{Q{\bar Q}\rightarrow \gamma+\gamma}$   of  an ordinary $Q{\bar Q}$ bound state into two photons.  From   Equation \eqref{eqn:230} it then follows that
\begin{align}\label{eqn:2321}
N_{Q{\bar Q}}^{V\!F}\sim \frac{1}{L_{Q{\bar Q}}^3}\times 2  \Gamma_{Q{\bar Q}\rightarrow \gamma+\gamma}\Delta t_{Q{\bar Q}} \sim 8c\left(\frac{M_{Q{\bar Q}}}{\hbar}\right)^2 \Gamma_{Q{\bar Q}\rightarrow \gamma+\gamma}\,.
\end{align}

Letting $m_Q$ and $q_Q$ denote, respectively, the mass and charge of the heavy quark $Q$ and letting $\omega_{Q{\bar Q}}^0$ denote the resonant angular frequency of the $Q{\bar Q}$ VF,  from  Equation \eqref{eqn:227} it the follows that the  contribution of the $Q{\bar Q}$ VF to $\epsilon_0$ is
\begin{equation}\label{eqn:2322}
{\epsilon_0}_{\left (\substack{\mbox{contribution}\\  \mbox{from $Q{\bar Q}$ VF}}\right)}\sim8c\left(\frac{M_{Q{\bar Q}}}{\hbar}\right)^2 \Gamma_{Q{\bar Q}\rightarrow \gamma+\gamma}
\frac{q_Q^2/(m_Q/2)}{(\omega_{Q{\bar Q}}^0)^2}\,.
\end{equation}

The least massive $c{\bar c}$ bound state, $\eta_c(1S)$, has $J=0$ and has positive charge conjugation parity\cite{Patrignani:16}, implying that a photon excited $\eta_c(1S)$ VF can decay into a single photon.  To obtain an order-of-magnitude estimate of  the maximum contribution that an $\eta_c(1S)$ VF could make to $\epsilon_0$, the experimental decay rate $\Gamma_{\eta_c(1S)\rightarrow \gamma+\gamma}$ of $\eta_c(1S)$ into two photons  is $7.69\times 10^{18}$/s\cite{Patrignani:16}.  It is not obvious which energy should be used to calculate the angular  frequency $\omega_{Q{\bar Q}}^0$,  but the minimum possible energy  $E_{\rm min}$  yields a minimum possible value for $\omega_{Q{\bar Q}}^0$ and a maximum possible value for the contribution of  an $\eta_c(1S)$ VFs to $\epsilon_0$.  $E_{\rm min}$ is the difference between the mass $m_{\eta_c(1S)}$ of $\eta_c(1S)$ and the masses $m_{\rm c}$ of the charm and anti-charm quarks when they are weakly bound: $E_{\rm min}=m_{\eta_c(1S)}-2m_{\rm c}\sim 2.98\,{\rm GeV}-2\times 1.27\, {\rm GeV}=0.44$ GeV\cite{Patrignani:16}.  From  Equation \eqref{eqn:2322} the maximum possible contribution to $\epsilon_0$ from $\eta_c(1S)$ VFs is then calculated to be 
\begin{equation}\label{eqn:2323}
{\epsilon_0}_{\left (\substack{\mbox{ maximum possible}\\ \\  \mbox{contribution from $\eta_c(1S)$ VFs}}\right)}\sim 1.3\times 10^{-3}\frac{e^2}{\hbar c}\,.
\end{equation}
Comparing  Equation \eqref{eqn:2323} with  Equation \eqref{eqn:232}, the maximum contribution to $\epsilon_0$ from $\eta_c(1S)$ VFs is about  $10^{-4}$ times smaller than the combined contribution from the three charged lepton-antilepton VFs.

 The  the least massive $b{\bar b}$ bound state, $\eta_b(1S)$, has $J=0$ and has positive charge conjugation parity\cite{Patrignani:16}. Its mass is known experimentally, but the decay rate into two photons is not\cite{Patrignani:16}. There are, however, theoretical calculations of the decay rate that range from 0.22 keV  to 0.45 keV\cite{Munz:96,Ebert:03,Kim:05,Huang:97}. To determine the maximum  contribution to $\epsilon_0$ from oscillations of $\eta_b(1S)$ VFs, the maximum value for the decay rate and the minimum value of energy associated with the state are used: $E_{\rm min}$ is the difference between the mass $m_{\eta_b(1S)}$ of $\eta_b(1S)$ and the masses $m_{\rm b}$ of the bottom and anti-bottom quarks when they are weakly bound: $E_{\rm min}=m_{\eta_b(1S)}-2m_{\rm b}\sim 9.40\,{\rm GeV}-2\times 4.3\, {\rm GeV}=0.8$ GeV\cite{Patrignani:16}. From  Equation \eqref{eqn:2322}  The maximum contribution to $\epsilon_0$ from oscillations of $\eta_b(1S)$ VFs is then calculated to be 
\begin{equation}\label{eqn:2324}
{\epsilon_0}_{\left (\substack{\mbox{ maximum possible}\\  \\ \mbox{ contribution from $\eta_b(1S)$ VFs}}\right)}\sim 2.6 \times 10^{-5}\frac{e^2}{\hbar c}\,,
\end{equation}
which is about  $10^{-6}$ times smaller than the contribution from the three charged lepton-antilepton VFs.
 
For the heavy quarks $c$ and $b$, as the mass increases from $m_c$ to $m_b$, the minimum possible angular frequency increases, and the decay rate of $Q{\bar Q}$ into two photons decreases. Both effects decrease the maximum possible contribution to  $\epsilon_0$ and suggest the contribution to $\epsilon_0$ from the top quark $t$ would be smaller than that from either $c$ or $b$.  On the other hand, the square of the charge for $t$  is four times larger than that for $b$.  Although there is no experimental information about $\eta_t(1S)$\cite{Patrignani:16}, from the above discussion the contribution of $\eta_t(1S)$ VFs to $\epsilon_0$ is expected to be small compared with the contribution from the three charged lepton-antilepton VFs.
 
 For the light quarks $q=u, d,\, {\rm or}\, s$, the $\pi^0, \eta$, and $\eta^\prime$  are the least massive $J=0$ combinations of $q{\bar q}$ bound states that decay into two photons. Here, however, the comparison with an oscillator is less appropriate; a Bethe-Saltpeter approach to $q{\bar q}$ bound states fails completely\cite{Gara:90}.  As discussed in Ref. \cite{Biernat:14}, a completely relativistic approach is required, an approach that shows no indication that a $q{\bar q}$ pair can be characterized by an oscillator potential energy.  Moreover, since the strong interactions are primarily responsible for the binding of these relativistic states,  $q{\bar q}$ bound state VFs would have much higher natural frequencies than the electromagnetically bound,  charged lepton-antilepton VFs that are much more weakly bound.  Accordingly $q{\bar q}$ bound state VFs  contribute little to $\epsilon_0$ and to lowest order need not be considered.
 
 \subsection{Calculation of $\epsilon_0$}
 \label{subsec:6.6}
 
Ignoring the small contributions to $\epsilon_0$ from quark-antiquark VFs, an approximate formula for  $\epsilon_0$  is immediately obtained from  Equation \eqref{eqn:232} using the defining formula $\alpha=e^2/(4 \pi \epsilon_0 \hbar c)$ to eliminate $\alpha$ and then using $c=1/\sqrt{\mu_0\epsilon_0}$: 
\begin{equation}\label{eqn:233}
\epsilon_0 \cong   \frac{ 6\mu_0}{\pi}\left(\frac{8e^2}{\hbar}\right)^2= 9.10\times 10^{-12}\rm \frac{C}{Vm}\, .
\end{equation}

The calculation of $\epsilon_0$ in  Equation \eqref{eqn:233} is strictly a quantum calculation: (1) The existence of on-mass-shell,  charged lepton-antilepton VFs are predicted by quantum field theory. (2) The appearance of charged lepton-antilepton VFs as  bound states in the lowest energy level with $J=0$ is predicted by the Heisenberg uncertainty principle and conservation of angular momentum. (3) The energy levels of the bound state VFs are calculated using non-relativistic quantum mechanics. (4) The decay rates of the charged  lepton-antilepton bound state VFs are calculated using quantum electrodynamics. (5) The calculation of $\epsilon_0$ from (1)-(4)  is a quantum derivation. Once  (1)-(4) are known, however, it is possible to obtain the formula  Equation \eqref{eqn:233} using a classical derivation similar to that used to obtain the permittivity of a physical dielectric\cite{Mainland:17a}.

 From   Equation \eqref{eqn:233}   equations for $c$ and $\alpha$ immediately follow as shown in the Introduction.  Only the lowest-order terms in $\alpha$ have been retained in calculating $\epsilon_0$.  The  binding energy of parapositronium was neglected when calculating $\Delta t_{p-Ps}$ in  Equation \eqref{eqn:17}, and only the leading term was retained when calculating $\omega_j^0$ and $\Gamma_{\rm p-Ps}$. The formulas for $\epsilon_0$, $c$, and $\alpha$ exhibit the point of view that $e$ and $\hbar$ are fundamental constants. (In SI units $\mu_0$ is a parameter with assigned a value.) The calculation of $\epsilon_0$ also describes how the properties of the vacuum determine the speed of light in the vacuum and the fine-structure constant.
 
\section*{References}

\bibliographystyle{iopart-num}
\bibliography{MainlandMulligan}
\end{document}